\documentclass[11pt]{scrartcl}

\usepackage[utf8]{inputenc}
\usepackage[english]{babel}

\usepackage[paper=a4paper,left=26mm,right=26mm,top=25mm,bottom=30mm]{geometry}

\usepackage{amsmath,amsfonts}
\usepackage{graphicx}
\usepackage{authblk}
\usepackage{booktabs}
\usepackage{array}
\newcolumntype{C}[1]{>{\centering\let\newline\\\arraybackslash\hspace{0pt}}m{#1}}

\usepackage{natbib}
\usepackage[breaklinks, colorlinks=true, citecolor=black, urlcolor=black, linkcolor=black]{hyperref}

\title{Comparison of Model Output Statistics and Neural Networks to Postprocess Wind Gusts}

\author[1]{Cristina Primo}
\author[2]{Benedikt Schulz}
\author[2]{Sebastian Lerch}
\author[1]{Reinhold Hess}
\affil[1]{Deutscher Wetterdienst}
\affil[2]{Karlsruhe Institute of Technology}

\date{\today}

\begin{document}

\maketitle

\begin{abstract}
\noindent
Wind gust prediction plays an important role in warning strategies of national meteorological services due to the high impact of its extreme values. However, forecasting wind gusts is challenging because they are influenced by small-scale processes and local characteristics. To account for the different sources of uncertainty, meteorological centers run ensembles of forecasts and derive probabilities of wind gusts exceeding a threshold. These probabilities often exhibit systematic errors and require postprocessing. Model Output Statistics (MOS) is a common operational postprocessing technique, although more modern methods such as neural network-bases approaches have shown promising results in research studies. The transition from research to operations requires an exhaustive comparison of both techniques. Taking a first step into this direction, our study presents a comparison of a postprocessing technique based on linear and logistic regression approaches with different neural network methods proposed in the literature to improve wind gust predictions, specifically distributional regression networks and Bernstein quantile networks. We further contribute to investigating optimal design choices for neural network-based postprocessing methods regarding changes of the numerical model in the training period, the use of persistence predictors, and the temporal composition of training datasets. The performance of the different techniques is compared in terms of calibration, accuracy, reliability and resolution based on case studies of wind gust forecasts from the operational weather model of the German weather service and observations from 170 weather stations.
\end{abstract}

\section{Introduction}
\label{sec:1}

Presently, the majority of weather forecasts rely on the outcomes generated by physical models, wherein systems of partial differential equations characterizing atmospheric processes are solved on a grid. In order to quantify forecast uncertainty and obtain probabilistic projections, these numerical weather prediction (NWP) models are run multiple times with diverse sets of initial conditions and/or model physics. This process yields an ensemble of predictions. Despite notable advancements in recent decades, ensemble forecasts still show systematic errors, necessitating correction through statistical postprocessing methods to obtain accurate and reliable forecasts \citep{VannitsemEtAl2021}.

Postprocessing has been a focus of research interest at the intersection of statistics and atmospheric sciences. Over the past decades, a large variety of methods has been developed and successfully applied in research and operations, see \citet{VannitsemEtAl2021} for an overview. A particular focus has been on probabilistic forecasts that provide information about forecast uncertainty, the estimation and communication of which is especially important for forecasts of hazardous weather conditions such as severe wind gusts.  Modern postprocessing methods correct systematic errors with distributional regression models for the conditional probability distribution of the weather variable of interest given predictors from the NWP system. In many of these postprocessing models, the probabilistic forecast is given by a parametric probability distribution with parameters depending on predictors through a suitably chosen link function \citep[e.g.\ ensemble model output statistics, EMOS,][]{Gneiting2005}.  Summary statistics of ensemble predictions of the variable of interest are often used as sole covariates. Over the past years, deep learning methods based on neural networks (NNs) that were first proposed in \citet{Rasp2018} have been demonstrated to provide state-of-the-art predictive performance and have seen broad application and extensions towards other variables \citep[e.g.,][]{Ghazvinian2021,Schulz2022,GneitingEtAl2023}, types of inputs \citep[e.g.,][]{Veldkamp2021wind, Chapman2022}, or time-scales \citep[e.g.,][]{scheuererBasisfunc,HoratLerch2023}.

Despite the promising research results, hardly any of the novel machine learning methods have been used in an operational context at meteorological services thus far, and a number of methodological and practical challenges remain for a successful transfer from research to operations \citep{VannitsemEtAl2021,Haupt2021}. During many years, the German weather service (\textit{Deutscher Wetterdienst}, DWD) has postprocessed forecasts by using model output statistics (MOS) techniques based on linear regressions. The currently operational technique, MOSMIX, as well as the weather warning guidance, WarnMOS, correct errors in deterministic outputs of a combination of two different global NWP systems: the Icosahedral Nonhydrostatic (ICON) Model at DWD and the Integrated Forecasting System (IFS) at the European Centre for Medium-Range Weather Forecasts (ECMWF). Both MOS-techniques consist of a two step process. In a first step, ICON and IFS are postprocessed individually providing ICON-MOS and IFS-MOS forecasts and in a second step, the resulting optimized forecasts are combined in the best possible way. MOSMIX currently postprocesses meteorological parameters commonly measured at observation sites, e.g., temperature and dew point in 2 m height, speed and direction of wind and wind gusts in 10 m height, expected precipitation amounts, precipitation phases (rain or snow), air pressure, sun shine duration, and many more. 
The postprocessing aims at providing optimized forecasts that are downscaled from the NWP model grid to user-relevant points of interest (5,400 locations worldwide). 
Further, WarnMOS is oriented to warning events and provides threshold exceedance probabilities for several meteorological variables.

In order to provide skillful probabilistic forecasts for weather warning events, a new system was implemented at DWD, ModelMIX \citep{Hirsch2014, Hess2020}, which not only combines the deterministic models, but also accounts for the uncertainty given by the ensemble prediction systems, ICON-EPS and IFS-EPS, as well as the uncertainty in convective permitting scales, the ensemble of the limited area model from the COnsortium for Small-scale MOdelling, COSMO-DE-EPS\citep{Baldauf2011}. All these systems are first individually postprocessed to obtain MOS forecasts, and then statistically combined in an optimal way to produce the probability of a warning event with uniform spatial and temporal resolution. The resulting probabilities are then internally used to produce automated warning proposals to support warning management at DWD.

In an attempt to modernize postprocessing at DWD with techniques that have shown promising results in research applications (e.g., NNs), efforts are in place now in assessing the benefits of using NN techniques in comparison with the current MOS approaches. It is important to remark that our work presented here focuses on wind gusts forecasts for the next 21 hours, whereas MOS techniques at DWD provide postprocessing of many more meteorological parameters (approx.\ 150) and up to 240 hours. Nevertheless, as a starting point in the comparison, we focus on postprocessing of one variable, which is particularly important for many warnings, to better understand the benefit of using new techniques, as well as to develop ways to account for potential issues in the transition from research to operations.
In addition, the underlying wind gust dataset has been used in previous research \citep{PantillonEtAl2018,Schulz2022,HoehleinEtAl2023}, and various skillful benchmark models for postprocessing are thus available.

The overarching aim of this article thus is to contribute to closing the gap between research on NN-based postprocessing methods and operational practices at weather services. Our contributions are twofold: On the one hand, we provide a comprehensive comparison of the forecast performance of NN models for postprocessing from research \citep{Schulz2022} to a pre-operational MOS technique for ensembles at DWD, ModelMIX\@. Hereby, we focus on probability forecasts for exceeding warning level thresholds of wind gust. 
On the other hand, we conduct experiments to investigate design choices of NN-based postprocessing models regarding their capabilities to cope with challenges that occur in the operational use of postprocessing models, such as changes in the NWP model. 
Changes to the resolution, model physics or components of the NWP system may invalidate the assumption of homogeneous error characteristics of the ensemble forecasts in the training data and thus represent a major challenge for the operational implementation of postprocessing models, in particular for complex models relying on the availability of large archives of past forecast and observation data for training. 
For EMOS models, \citet{LangEtAl2020} find that utilizing multiple years of training data irrespective of NWP model changes during that period show superior performance compared to time-adaptive training schemes based on sliding windows where only forecast cases from the most recent days are considered, even though the latter should be able to more quickly adapt to changes in forecast error characteristics.
However, it is an open question how changes in the NWP system affect the forecast performance of NN-based postprocessing models and how those can best be adapted to deal with such updates. 
In addition to the temporal composition of the training datasets, we further investigate the effect of other design choices for NN-based postprocessing models motivated by analogous techniques used in operational practice, specifically the incorporation of persistence-based predictor variables, and the joint modeling of multiple lead times.

We present two case studies utilizing forecasts from the operational EPS of DWD, which underwent several significant updates since its inception in 2010. Our focus is on probabilistic forecasts of exceedances of warning level thresholds for wind gusts, and we compare different variants of statistical and NN-based postprocessing models to address the research questions outlined above.

The remainder of this article is organized as follows. Section \ref{sec:data} introduces the dataset used in the case studies, and Section \ref{sec:pp} presents the postprocessing models. Following a short overview of verification methods in Section \ref{sec:4}, the main results are presented in Section \ref{sec:5}. The article closes with a discussion in Section \ref{sec:conclusions}.

\section{Data}
\label{sec:data}

\subsection{NWP Model Data and Observations}
\label{subsec:data_nwp}

\begin{figure}
	\centering
	\includegraphics[width=0.99\textwidth]{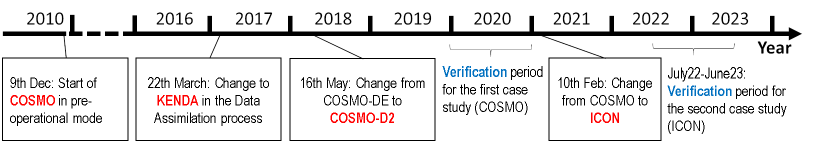}
	\caption{Overview of relevant changes in the NWP model during the time range of the data considered.}
	\label{fig:Evolution}       
\end{figure}

The forecasts are generated by the operational 20-member EPS at convection-permitting resolution of DWD. The EPS has undergone several significant changes over time, which are specified in the following and summarized in Fig.\ \ref{fig:Evolution}. 

The EPS started as the COSMO-DE-EPS in pre-operational mode in December 2010. The ensemble members are obtained from physically perturbing the initial and boundary conditions of four different global models combined with five sets of physical perturbations \citep{Gebhardt2008,Peralta2012}. The model runs at a horizontal grid spacing of 2.8km and is initialized every 3h, with hourly forecasts produced up to 21h ahead. 
In 2017, the lateral boundary conditions started being driven by the global ICON-EPS system, and initial conditions were given by a kilometer-scale ensemble data assimilation system \citep[KENDA,][]{Schraff2016}. In May 2018, the model was extended to better represent convective structures, with COSMO-DE being replaced by COSMO-D2, where the horizontal and vertical resolution increased, as well as the horizontal area covered.

In February 2021, the convection-permitting model setup ICON-D2 (i.e., using the limited-area mode of the new NWP system ICON) replaced COSMO-D2 \citep{Reinert2023}. Starting values of ICON-D2-EPS are based on the local ensemble transform kalman filter (LETKF) data assimilation of the global model ICON-EPS. Its grid consists of a set of spherical triangles that allows for a nearly constant grid spacing. 
Our first case study utilizes data from 2020 for verification and focuses on COSMO-D2-EPS, whereas the second case study is based on verification data from July 2022 to June 2023 and the ICON-D2-EPS model.

\begin{figure}
\centering
\includegraphics[width=0.4\textwidth]{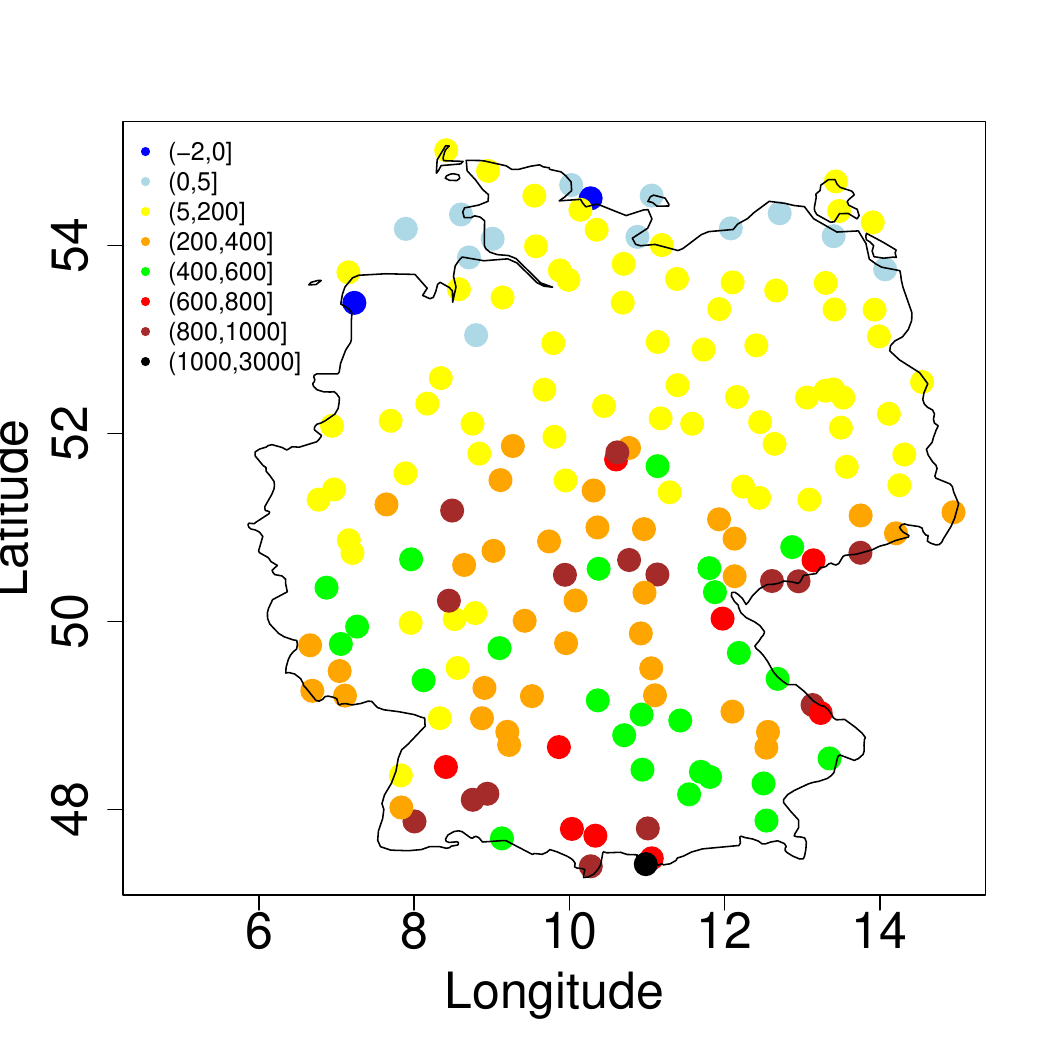}
\caption{Map of SYNOP stations. The color represents the station height (m).}
\label{fig:Stats}       % Give a unique label
\end{figure}

Observations come from 170 SYNOP stations in Germany operated by DWD, which are displayed in Fig.\ \ref{fig:Stats}. Each station is paired with the nearest grid point to create pairs of observations and forecasts.

\subsection{Predictors and Target}
\label{subsec:data_preds}

In this work, we use the same set of predictors for ModelMIX as in \citet{Hess2020} and the same for the NNs as in \citet[see their Table 1 for details]{Schulz2022}. Those predictors include basically all meteorological variables that the model generates as well as selected temporal and spatial information.
In addition, some of the postprocessing methods presented in Section \ref{sec:pp} make use of temporal predictors that encode the model version. For each major model update (KENDA, D2, ICON), we introduce a binary variable that jumps at the time of the model change.
Finally, we briefly comment on persistence predictors, which are used by some of the postprocessing methods evaluated in Section \ref{sec:5}. Here, the persistence predictors refer to the most recently observed wind gust values at the associated station, that is, at the time the NWP model is initialized as well as one and two hours later.
As discussed below, our evaluation starts at a lead time of three hours to mimic an operational setup, and incorporating persistence predictors for the first two hours thus is admissible.

Our interest lies in predicting the threshold exceedance probabilities of hourly wind gusts for a particular list of thresholds the DWD uses, namely 25, 27, 33, 40, 47, 55, 63 and 75kt. In this work, we mostly focus on the first and lowest threshold, which is most frequently exceeded. The NWP model provides an ensemble of wind gusts as maximum wind speed values over the last hour. Of the available NWP model runs, we will focus on those initialized at 00UTC and 12UTC, with hourly outputs from 3h up to 21h. It takes around two hours to have all forecasts from the reference NWP (COSMO-DE-EPS, COSMO-D2-EPS or ICON-EPS) available and any postprocessing method used operationally could only run after having these data. Therefore, the first two lead times (1h and 2h) are not considered here.

\section{Postprocessing Methods}
\label{sec:pp}

In the following, all postprocessing methods and variants thereof, which are considered in the comparisons, will be introduced. First, we present ModelMIX, the postprocessing technique used at DWD, then we present the approaches from research on probabilistic wind gust forecasting.

An overview is available in Table \ref{tbl:pp_methods}, divided by the two case studies. Note that the computing times stated in the table should only be seen as rough approximations and are not directly comparable due to differences in the underlying hardware and software environments used for model estimation.

\begin{table}[p]
    \begin{center} 
    %		\small
    \scalebox{0.8}{
        \begin{tabular}{>{}l@{\hspace{2em}}c@{\hspace{2em}}c@{\hspace{2em}}c@{\hspace{2em}}l@{\hspace{2em}}l@{\hspace{2em}}r@{\hspace{2em}}r} 
        	\toprule
            Method & \multicolumn{3}{c}{Estimation and predictors} & \multicolumn{3}{c}{Training set} & Runtime  \\ 
            \cmidrule(r){1-1} \cmidrule(r){2-4} \cmidrule(r){5-7} \cmidrule(r){8-8}
        	Abbreviation & Local & Joint & Persistence & Period & Start & Samples & Total \\
        	\midrule
            \multicolumn{8}{l}{\textit{Training until December 2019, verification on COSMO-D2-EPS until December 2020}} \\
            ModelMIX$_1$ & (\checkmark) & (\checkmark) & \checkmark & COSMO & 09/12/2010 & **500,000 & **48 h \\
        	EMOS & \checkmark & - & - & Post-D2 & 16/05/2018 & 585 & 3.13 min \\
        	EMOS-GB & \checkmark & - & - & COSMO & 09/12/2010 & 3,243 & 34.18 h \\
        	BQN$_{1, \text{C}}$ & - & - & - & COSMO & 09/12/2010 & 562,462 & 39.34 h \\
        	BQN$_{1, \text{D}}$ & - & - & - & Post-D2 & 16/05/2018 & 101,749 & *6.78 h \\
        	DRN$_{1, \text{C}}$ & - & - & - & COSMO & 09/12/2010 & 564,217 & 41.97 h \\
        	DRN$_{1, \text{D}}$ & - & - & - & Post-D2 & 16/05/2018 & 101,749 & *7.60 h \\
        	DRN$_{1, \text{C}}^{\text{P}}$ & - & - & \checkmark & COSMO & 09/12/2010 & 564,217 & 41.21 h \\
        	\multicolumn{8}{l}{} \\
            \multicolumn{8}{l}{\textit{Training until June 2022, verification on ICON-D2-EPS until July 2023}} \\
            ModelMIX$_2$ & (\checkmark) & (\checkmark) & \checkmark & COSMO+ICON & 09/12/2010 & **639,000 & **48 h \\
        	DRN$_{2, \text{I}}^{\text{P}}$ & - & - & \checkmark & ICON & 10/02/2021 & 82,701 & 5.59 h \\
            DRN$_{2, \text{K}}^{\text{P}}$ & - & - & \checkmark & Post-KENDA & 22/03/2017 & 323,675 & 24.16 h \\
            DRN$_{2, \text{I}}^{\text{P, J}}$ & - & \checkmark & \checkmark & ICON & 10/02/2021 & 3,142,654 & 6.15 h \\
            DRN$_{2, \text{K}}^{\text{P, J}}$ & - & \checkmark & \checkmark & Post-KENDA & 22/03/2017 & 12,299,665 & 48.39 h \\
        	\bottomrule 
        \end{tabular}
        }
    \end{center}
    \caption{Overview of the main characteristics of the different postprocessing methods/variants. 
    The bottom-index of the methods refers to the verification period considered (1 = 2020, 2 = 2022/2023) and the training period (C = COSMO, D = Post-D2, K = Post-Kenda, I = ICON), the top-index refers to whether persistence predictors were used (P = including persistence) and whether a joint model for all lead times and initializations was estimated (J = joint model).
    The column 
    '\textit{samples}' refers to the (average) training set size for each individual model, '\textit{total runtime}' refers to the total time required to train and predict for all models. The total run time of BQN$_{1, \text{D}}$ and DRN$_{1, \text{D}}$ is marked with a star (*), as the model was run for only one of the two initializations. For comparison with the other variants, we double this runtime.
    ModelMIX is also marked (**) since it postprocesses simultaneously about 150 variables in a parallelized process, so provided numbers are estimated values from the total sample size and total run time comprising all variables.} 
    \label{tbl:pp_methods}	
\end{table}

\subsection{Model Output Statistics (MOS)}
\label{subsec:pp_mos}

ModelMIX is a classical MOS technique and is addressed here as a reference method. It is based on an operational MOS scheme at DWD that postprocesses the deterministic forecasts of ICON-D2 and IFS of ECMWF. ModelMIX has been enhanced for the ensemble forecasts of COSMO-DE/D2-EPS and ICON-D2-EPS and is tailored for extreme events of almost all relevant weather warnings of DWD.

Threshold exceedance probabilities are generated in two steps: First, wind gust velocities are optimized using multiple linear regressions. Ensemble mean and spread of 56 model variables are used as model input, from which altogether about 300 predictors are derived including area means of model variables around the locations for which training is performed.

Each hourly time step is trained separately and for thresholds up to 27kt individually for each location. For thresholds above 33kt, locations with similar climatology are aggregated in 9 clusters and training is performed for each cluster in order to gather enough events for statistical modeling.

Based on the wind gust velocities, global logistic regressions are fitted for all locations together and parameterized for all time steps, see \citet{Hess2020} for more details \citep[Note that][refers to ModelMIX as Ensemble-MOS]{Hess2020}. Location-specific characteristics are considered to be taken into account already by the individual optimization of the velocities. The logistic parameterization focuses on the time step dependent forecast errors and allows for providing meaningful statistical forecast for high thresholds up to 75kt, where the number of events is too small for training for individual stations or even time steps.

For the first case study (verification period 2020), training was performed using forecasts of COSMO-DE-EPS and COSMO-D2-EPS together up to the end of 2019. For the second case study (July 2022 to June 2023), additionally forecasts of ICON-D2-EPS were used until June 2022. 
The model change from COSMO-D2-EPS to ICON-D2-EPS is taken into account by the binary model update predictor and by double weighting of ICON-D2-EPS in the training data set.

Even though ModelMIX does not run operationally, it was designed for an operational time lapse, therefore it starts to process forecasts after a time step of 3h. Moreover, the latest observations at 2h after nominal model issue are available at start of the postprocessing and are used as persistence predictors in order to boost the performance of short-term statistical forecasts.

The computing time of ModelMIX is about 15min including preparation of numerical and observation data as input. The evaluation of the statistical model itself is computed in parallel on DWD's high performance computer and requires less than 2min for an issue of ICON-D2-EPS comprising about 10$^6$ statistical equations altogether (predicting about 150 statistical forecasts for almost all weather warnings of DWD).

\subsection{Postprocessing methods leading to predictive distributions}
\label{subsec:pp_nn}

While ModelMIX generates probability forecasts tailored to the specific thresholds,
another approach for the generation of threshold exceedance probabilities is to first develop a model for the entire distribution of the target variable and to then derive the desired probabilities from that distribution. This approach has not only the advantage that probabilities for any threshold can be calculated, but also that, if multiple thresholds are considered, only one model needs to be estimated. 
Further, this guarantees that threshold exceedance probabilities are decreasing with respect to the threshold.
On the other hand, models specifically tailored to certain thresholds will likely result in superior performance, as they directly target the specific event probability of interest during estimation.

Much work in research on statistical postprocessing of ensemble forecasts has focused on forecasting the entire distribution of the target variable \citep[see, e.g.,][]{Vannitsem2018}, and a variety of methods for that task is readily available. \citet{Schulz2022} provide a systematic comparison of ensemble postprocessing methods for wind gusts and find that NN-based postprocessing methods outperform established state-of-the-art postprocessing approaches. Based on their work, we apply the NN methods to generate threshold exceedance probabilities. In addition, we also use two of the benchmark methods of this study for a broader comparison.

As \citet{Schulz2022} apply the methods to the same NWP model and observational database (although for the period 2010--2016), we directly adopt their setup of the postprocessing methods. This includes the implementation details of the postprocessing methods, the choice of hyperparameters and that of the predictor variables. Note that we did not tune the hyperparameters specifically to the application at hand.
Hence, we refer to \citet{Schulz2022} for a detailed description of the postprocessing methods and accompanying code, and only describe changes with respect to this setup in the following.

\subsubsection{Ensemble Model Output Statistics (EMOS)}
\label{ssubsec:pp_emos}

Before introducing the NN methods, we briefly comment on the two benchmark methods considered in the comparison. The EMOS \citep{Gneiting2005} approach assumes that the target variable follows a parametric distribution, here, a truncated logistic distribution, and links the parameters of the distribution to summary statistics of the ensemble forecasts, where the link function is estimated minimizing a proper scoring rule. For the incorporation of additional predictor variables, the classic EMOS approach can be extended using ideas from machine learning, namely, gradient boosting (EMOS-GB) \citep{Messner2017}, which has shown to produce more accurate forecasts \citep{Rasp2018,Schulz2022}. 

Here, we follow the standard approach of estimating both variants locally and for each lead time separately. 
Since preliminary tests showed that using only data since the last model update was superior to a longer training period including binary predictors on the model version, we decided to use the former training scheme for the EMOS model.
This reduces the amount of training data such that we do not utilize the seasonal training approach of \citet{Schulz2022}.
For the gradient boosting extension of EMOS, incorporating the binary predictors on the model version is feasible. Therefore, we include those and use the entire COSMO period for training.

\subsubsection{Neural Network methods}
\label{ssubsec:pp_drn}

The distributional regression network (DRN) approach \citep{Rasp2018} extends the EMOS framework towards the use of methods from artificial intelligence, in that a NN links the predictor variables with the distribution parameters. Analogous to the EMOS framework, we employ a truncated logistic distribution as forecast distribution. 
NNs are data-driven models that typically work the best the more data is available to feed the model. For station-based postprocessing, similar observations have been made, in that a single model is estimated jointly for all stations, but made locally adaptive via a technique originally proposed in natural language processing, so-called embeddings, which notably improves the predictive performance. For details and comparisons to alterative setups, see \citet{Rasp2018}.

The change from ICON to COSMO represents a substantial change in the NWP model and, as only one year of training data for the prediction in 2022/2023 is available, the question of the ideal composition of the training set arises. While \citet{Schulz2022} fit a separate model for each lead time (for a fixed model run), we investigate, in case of the ICON data, the performance of a joint model for all lead times and the two model runs at 00 UTC and 12 UTC, thereby increasing the training set size by a factor of 38.
For the joint model, we change the target variable from the variable itself to the bias of the ensemble mean. This centers the target variable and is intended to help the model handle changes in the diurnal cycle, which is not present in the data when considering each lead time separately, as the forecasts then all validate at the same time of the day. Further, we also standardize the target variable in this case (for training). Note that we now use a non-truncated logistic distribution for estimation and do not account for a possibly non-positive support of the forecast distribution of the wind gusts. The forecast, however, which is the sum of the ensemble mean and the forecast distribution of the bias, is truncated in zero.

While the joint model is only considered for the ICON data in 2022/2023, one general modification with respect to the setup in \citet{Schulz2022} concerns all NN models. Instead of using the last year within the training period for validation, we sample randomly 20\% of the training data to obtain a validation subset. We changed the selection in order to have a balanced portion of all model versions both within the data sets used for estimating and validating the NNs.

For predicting exceedance probabilities at high thresholds, the type of forecast distribution might be an influential factor, hence we consider one of the two more flexible alternatives to DRN presented in \citet{Schulz2022}. The Bernstein Quantile Network \citep[BQN;][]{Bremnes2020} approach differs from DRN in the choice of the forecast distribution and the associated loss function. BQN models the quantile function as a linear combination of Bernstein basis polynomials, a class of flexible function approximators. Note that the change in the DRN model regarding the choice of the validation set also applies to the BQN models.

\section{Verification Methods}
\label{sec:4}

Forecast quality has many different attributes such as accuracy, reliability, resolution or sharpness, but no verification measure is able to summarize all of them.
In this paper, we use standard tools tailored to the evaluation of probability forecasts, which we present in the following.
Reliability and sharpness diagrams are used to account for calibration and sharpness, and the Brier score (with its decomposition into three terms) to analyze the accuracy. 

\subsection{Reliability Diagrams}
\label{subsec:41}

The reliability diagram \citep{Sanders1963} is a graphical tool to assess whether the postprocessing methods appropriately quantify the uncertainty, i.e., if they are calibrated. In a calibrated system, forecast probabilities correspond to the observed frequencies. To construct the reliability diagram, forecasts probabilities of a selected event (a threshold being exceeded, in our case) are distributed into bins, and compared to the observed frequency for each bin. A perfectly calibrated system would lie on the diagonal, and the deviation from the diagonal gives the conditional bias depending on the forecast probability. A line close to the horizontal line given by the climatology indicates that the system is not able to distinguish situations with different frequencies of occurrence (no resolution). A system is skillful if the reliability diagram is closer to the diagonal than the intermediate line that lies between the diagonal and the climatology. 

On the other hand, each point of the reliability diagram refers to a different number of cases. Since we are dealing with extreme events which are comparatively rare, small probability forecasts tend to be much more frequent than higher values. 
The sharpness diagram depicts the relative frequency of the event in each forecast probability bin. In the case of a perfectly sharp probabilistic system, the sharpness diagram would consist of a histogram with two bars, one over the value zero for those cases when the model is sure no event would happen, and another one over the value one, when the model forecast the event. 
Sharpness is a property of the forecasts only, and -- subject to calibration -- sharper forecasts are preferable \citep{Gneiting2007}.

\subsection{Brier Score Decomposition}
\label{subsec:42}

To measure the accuracy, i.e., the level of agreement between probabilities and the binary observation, we use the Brier Score \citep[BS;][]{Brier1950}, defined as the mean squared error of the probability forecast. In a set of $n$ forecast-observation pairs, each probability forecast $f_i$ is compared to the binary observation $o_i$, that is 1 if the event occurs (i.e., the wind is above the threshold), and 0 otherwise,
\begin{equation}
BS=\frac{1}{n}\sum_{i=1}^n{(f_i-o_i)^2}.
\end{equation}
The BS always takes values between 0 and 1, and it can be decomposed into three terms, namely, the reliability, the resolution and the uncertainty term:
\begin{equation}
BS = \text{Rel}-\text{Res}+\text{Unc}.
\end{equation}
The reliability term refers to the agreement between forecast probability and mean observed frequency, which is also visualized in the reliability diagrams. The resolution shows if the observed outcome changes as the forecast changes. The uncertainty only depends on the variability of the observations. Hence, one should not compare the BS of events with different frequencies, because by definition, the lower the frequency of the event, the lower the contribution of the uncertainty term to the total BS, without necessarily implying a better accuracy. 

To analyze the degree of improvement of the postprocessed forecasts with respect to a reference system (with BS $BS_{ref}$), we can calculate the Brier skill score (BSS) via
\begin{equation}
BSS = \frac{BS_{ref}-BS}{BS_{ref}}. \label{eq:bss}
\end{equation}
If the BSS is negative, the reference system is superior to the postprocessed forecasts, if it is equal to zero, they have the same skill, and if it is positive the postprocessed forecasts outperform the reference with a maximum skill score of 1 indicating optimal predictive performance.

In addition, we can specifically analyze the improvement in terms of the resolution term. While the optimal BS is equal to 0 (together with a reliability term of 0), the associated resolution term is equal to the uncertainty term. Hence, the (positively oriented) resolution term is bounded by the uncertainty term, which we need to take into account when calculating a skill score analogous to the BSS.
To do this, we use the difference of the uncertainty term and the resolution term as underlying score in Eq.\ \eqref{eq:bss}, that is,
\begin{equation}
\frac{(\text{Unc} - \text{Res}_{ref}) - (\text{Unc} - \text{Res})}{(\text{Unc} - \text{Res}_{ref})} 
% = \frac{\text{Res} - \text{Res}_{ref}}{\text{Unc} - \text{Res}_{ref}} 
= \frac{\text{Res}_{ref} - \text{Res}}{\text{Res}_{ref} - \text{Unc} }. \label{eq:res_skill}
\end{equation}
This term can be interpreted analogously to the BSS. Note that the uncertainty component is identical for all forecasting systems.

\section{Results}
\label{sec:5}

This section shows a comparison of the different postprocessing techniques, namely (ensemble) MOS techniques and NNs. Since in an operational weather center, improvements in the initial conditions of the models, as well as in the numerical model itself, are frequently introduced, the relevance of these changes within the training period of the NNs is analyzed and discussed. For example, this includes the major changes in the data assimilation process of DWD (introduction of KENDA and later LETKF) and the change from COSMO-D2 to ICON-D2. 
This section also presents results regarding other important aspects in an operational center, e.g., ensuring the timely generation of postprocessed forecasts as well as design choices regarding the training of the networks.

\subsection{Comparison of MOS and NN methods}
\label{subsec:51}

We start with the first case study, where we compare the five different postprocessing methods introduced in Section \ref{sec:pp} with the direct model output of COSMO-D2-EPS.
Note that all model abbreviations in the plots are listed in Table \ref{tbl:pp_methods}.

\begin{figure}[t]
\includegraphics[width=\textwidth]{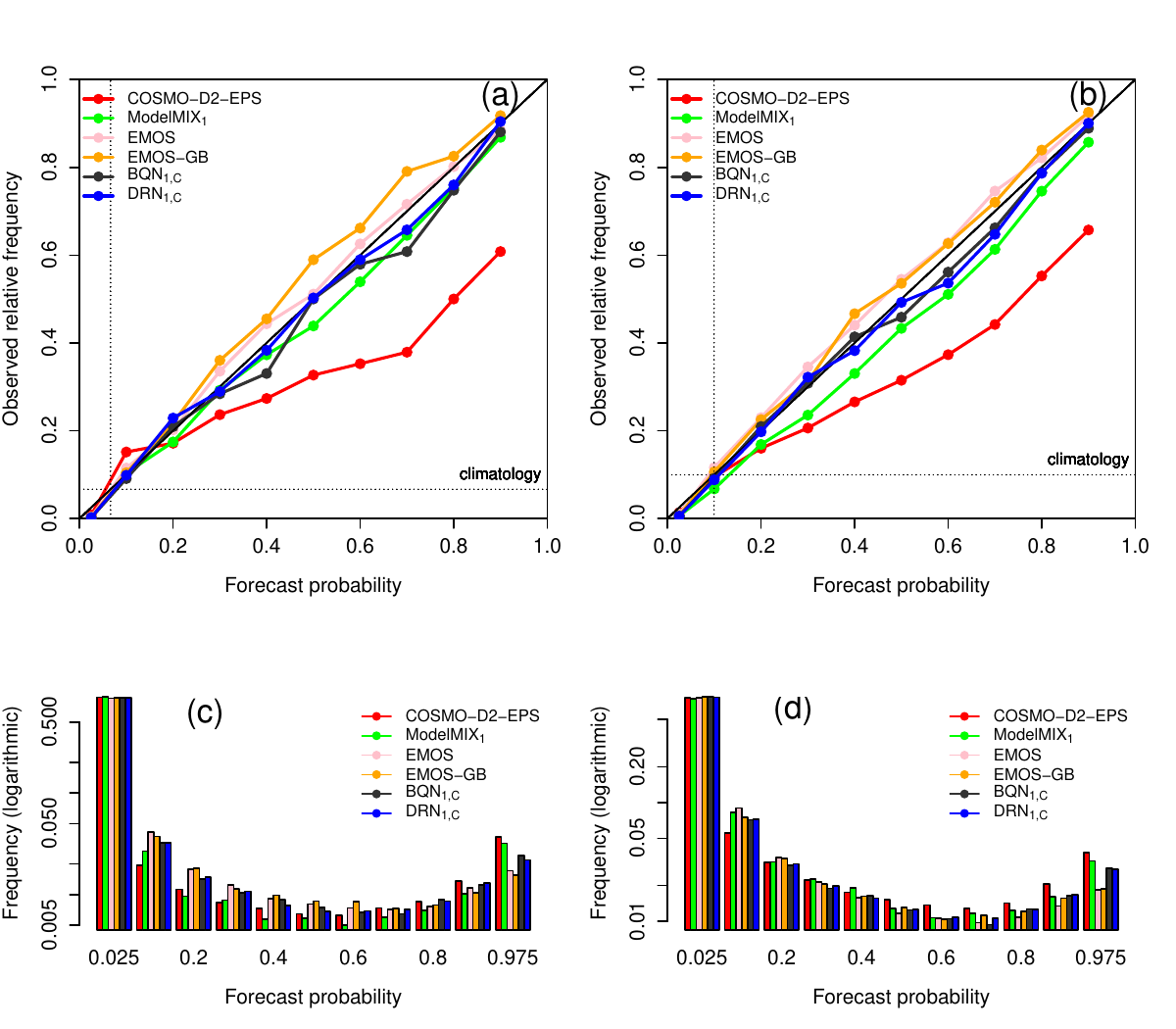}
\caption{Reliability diagrams (a-b) and sharpness histograms (c-d) of 
the five postprocessing techniques and the direct model output of COSMO-D2-EPS, for two different lead times: 3h (left) and 15h (right). The evaluation is based on data from year 2020, wind threshold 25kt, and the model run at 00UTC.}
\label{fig:ReldiagHist_Models}       % Give a unique label
\end{figure}

To assess calibration, Fig.\ \ref{fig:ReldiagHist_Models} shows the reliability diagrams of hourly wind gusts exceeding 25kt, for the model run at run 00UTC, for two different lead times, 3h (a) and 15h (b), as well as the corresponding sharpness diagrams (c and d, respectively). All postprocessing methods calibrate the direct model output and lie close to the diagonal, although EMOS and EMOS-GB tend to lie above the diagonal (underforecasting) whereas the other models below (overforecasting), especially ModelMIX for longer lead times.

The histograms in the sharpness diagrams are dominated by values close to zero, as expected. For small lead times, COSMO-D2 and ModelMIX are sharper (with more zero and one values than the other models) but this changes for higher lead times. Classic EMOS shows the lowest sharpness. 

\begin{figure}[p]
\includegraphics[width=\textwidth]{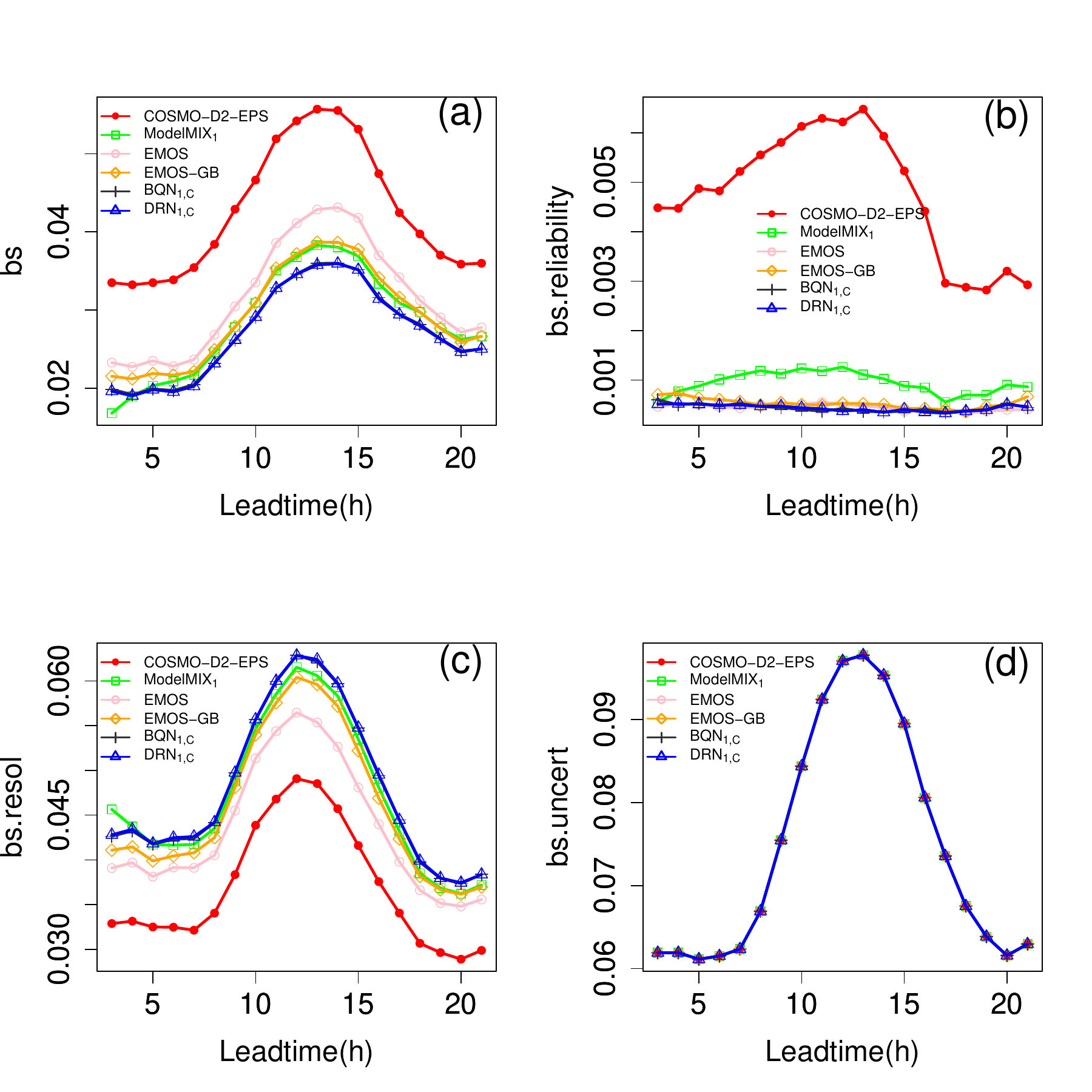}
\caption{Temporal evolution of the BS (a) and its reliability (b), resolution (c) and uncertainty (d) terms for the different postprocessing methods dependent on the lead time. The evaluation is based on data from year 2020, wind threshold 25kt, and the model run at 00UTC.}
\label{fig:BS_Models}       % Give a unique label
\end{figure}

Regarding accuracy, Fig.\ \ref{fig:BS_Models} shows the BS (a) and its decomposition in reliability (b), resolution (c) and uncertainty (d) terms. Postprocessing improves accuracy, and apart from the first few hours, where ModelMIX shows the highest accuracy, the NN methods show the best results. 
In line with previous research results \citep[e.g.,][]{Schulz2022} and other comparisons not shown here, there are only very minor differences between the two NN methods, DRN and BQN. This holds in particular also for high thresholds. We will therefore focus on DRN in the following. 
ModelMIX exhibits a notably worse reliability term, but the contribution of this term to the total BS is small. The improvement in the BS of NN compared to ModelMIX is mainly due to an improvement in resolution. Since all methods are verified against the same observations, the uncertainty term is equal for all of them. 

\begin{figure}[t]
\includegraphics[width=\textwidth]{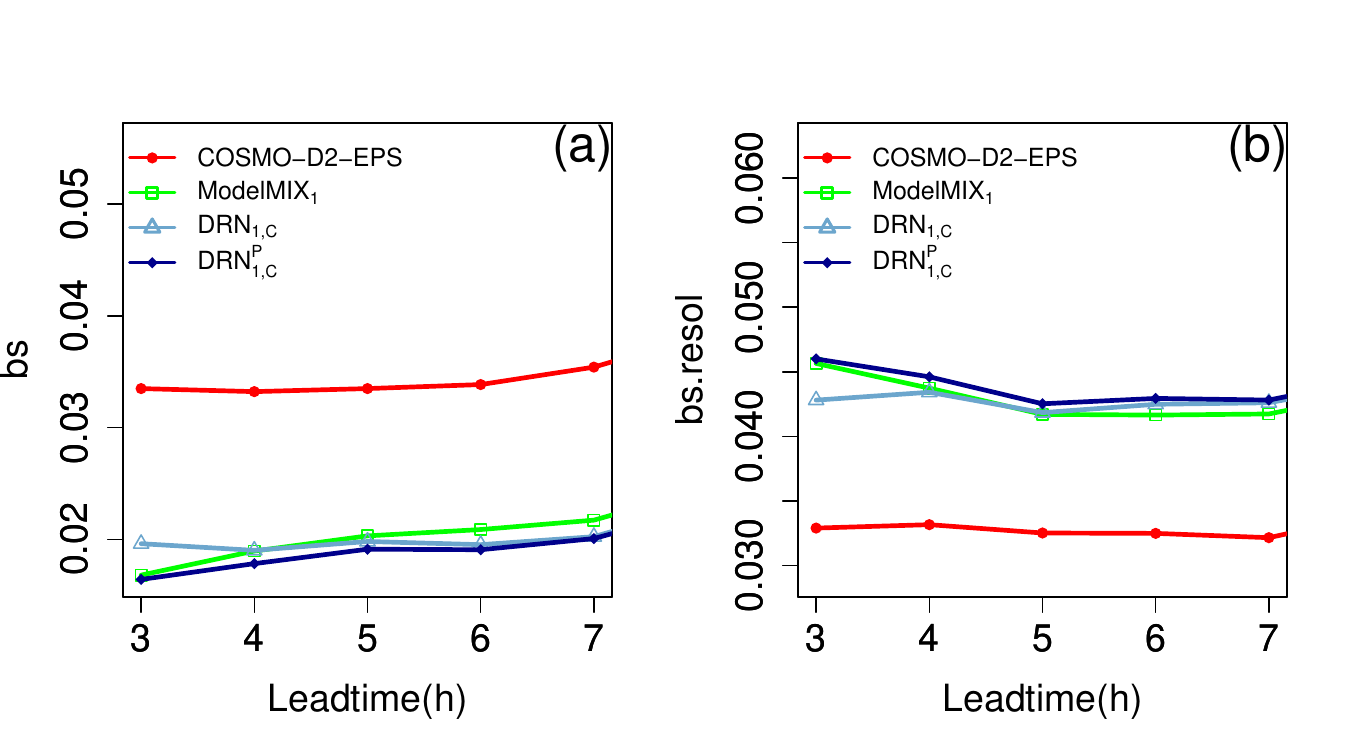}
\caption{Temporal evolution of (a) the BS and (b) the resolution term during the first hours when using persistence predictors. The evaluation is based on data from year 2020, wind threshold 25kt, and the model run at 00UTC.}
\label{fig:Persist}     
\end{figure}

The main difference regarding the predictors used by ModelMIX and the NNs, is that the NNs do not use any persistence predictors, whereas ModelMIX uses all available observations up to the time it is run operationally, i.e., two hours later than the run of the NWP model. To evaluate whether this effects the forecast performance during the first forecast hours, we run an experiment where DRN was trained using three additional persistence predictors: the wind gust observation at initialization time (of the NWP model), at initialization time plus 1h and at initialization time plus 2h. This setup mimics the use of persistence predictors by ModelMIX. Fig.\ \ref{fig:Persist} shows a comparison of previous models before and after using persistence predictors. These predictors notably improve the BS within the first hours, and the NN model with persistence predictors now slightly outperforms ModelMIX. In particular, the improvement is related to an increase in the resolution term.
Therefore, we will keep this setup for all further comparisons below.

\begin{figure}[t]
\includegraphics[width=\textwidth]{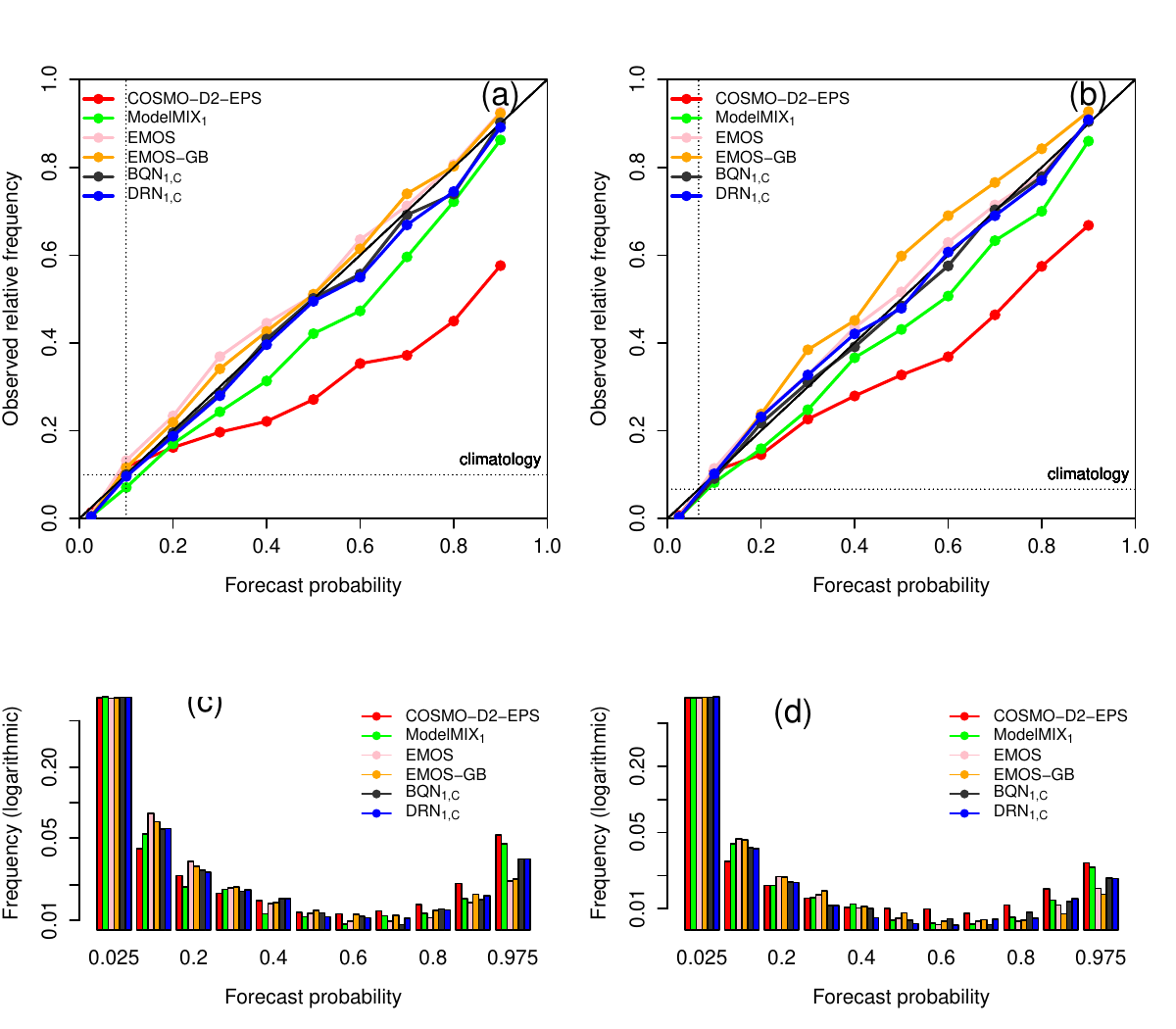}
\caption{Reliability diagrams (a-b) and sharpness histograms in logarithmic scale (c-d) of the five postprocessing techniques and the direct model output of COSMO-D2-EPS, for two different lead times: 3h (left) and 15h (right). The evaluation is based on data from year 2020, wind threshold 25kt, and the model run at 12UTC.}
\label{fig:ReldiagHist_Models12}       % Give a unique label
\end{figure}

\begin{figure}
\includegraphics[width=\textwidth]{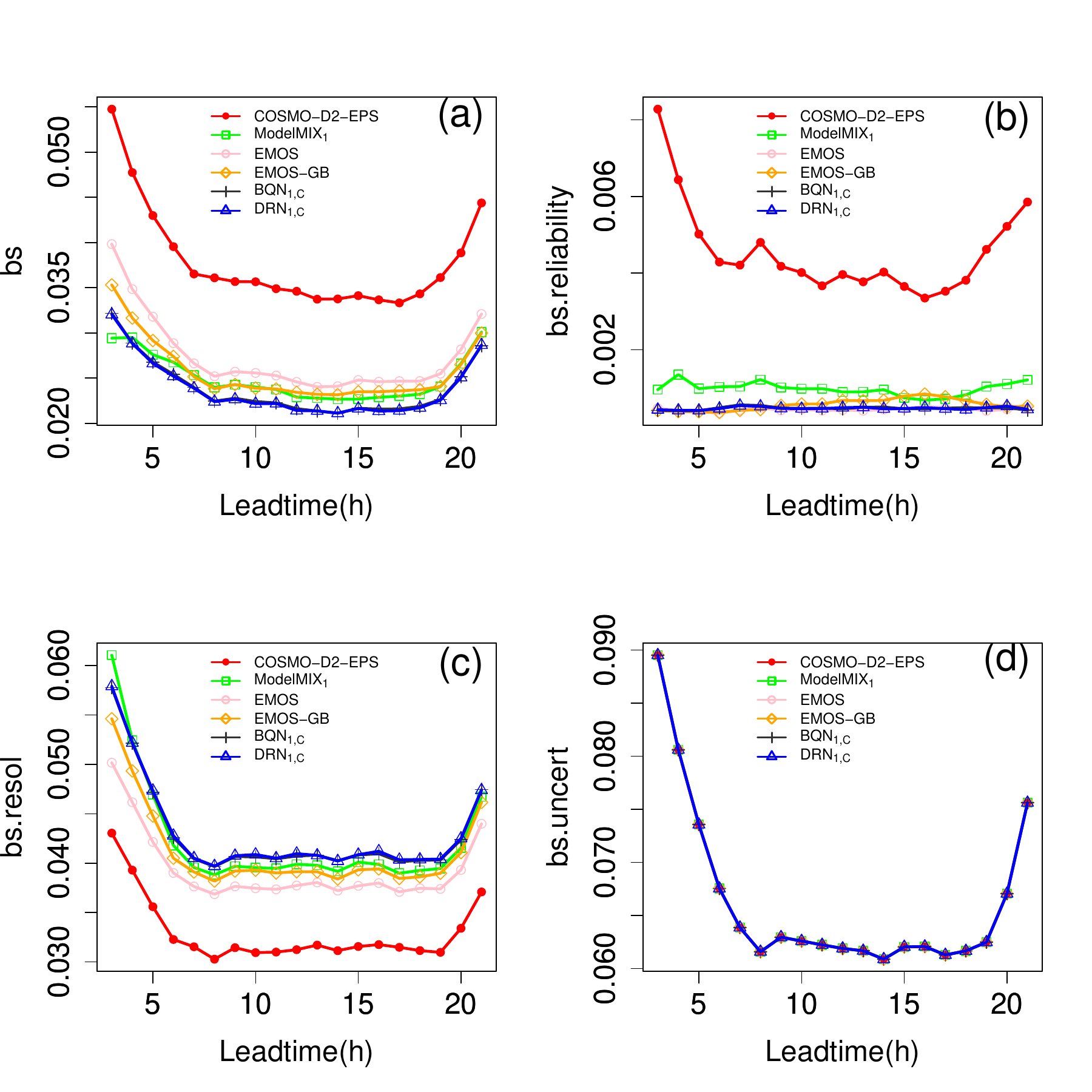}
\caption{Temporal evolution of the BS (a) and its reliability (b), resolution (c) and uncertainty (d) terms for different postprocessing methods dependent on the lead time. The evaluation is based on data from year 2020, wind threshold 25kt, and the model run at 12UTC.}
\label{fig:BS_Models12}       % Give a unique label
\end{figure}

Since the model performance is usually affected by the daily cycle, we further analyzed raw and postprocessed forecasts from the model run initialized at 12UTC to account for differences between day and night.
Fig.\ \ref{fig:ReldiagHist_Models12} shows the reliability diagrams and sharpness histograms of hourly wind gusts exceeding 25kt for lead time 3h (a;c) and 21h (b;d), respectively. 
COSMO-D2-EPS notably less well calibrated in this run, as well as ModelMIX. On the contrary, the EMOS techniques are better calibrated than for the model run initialized at 00UTC. 
Regarding accuracy, for which results are shown in Fig.\ \ref{fig:BS_Models12}, both runs show similar results. The EMOS techniques improve over the direct model output, and the NN methods outperform EMOS, with the improvements mainly being due to an improvement in resolution. 
The event of interest, i.e., the wind gusts exceeding 25kt, is observed mainly during daytime hours, and on those hours, forecasts show higher BS than during nighttime hours, where the event occurs less frequently. Therefore a direct comparison of BS for run 00UTC and run 12UTC should be avoided.

\begin{figure}
\centering 
\includegraphics[width=0.5\textwidth]{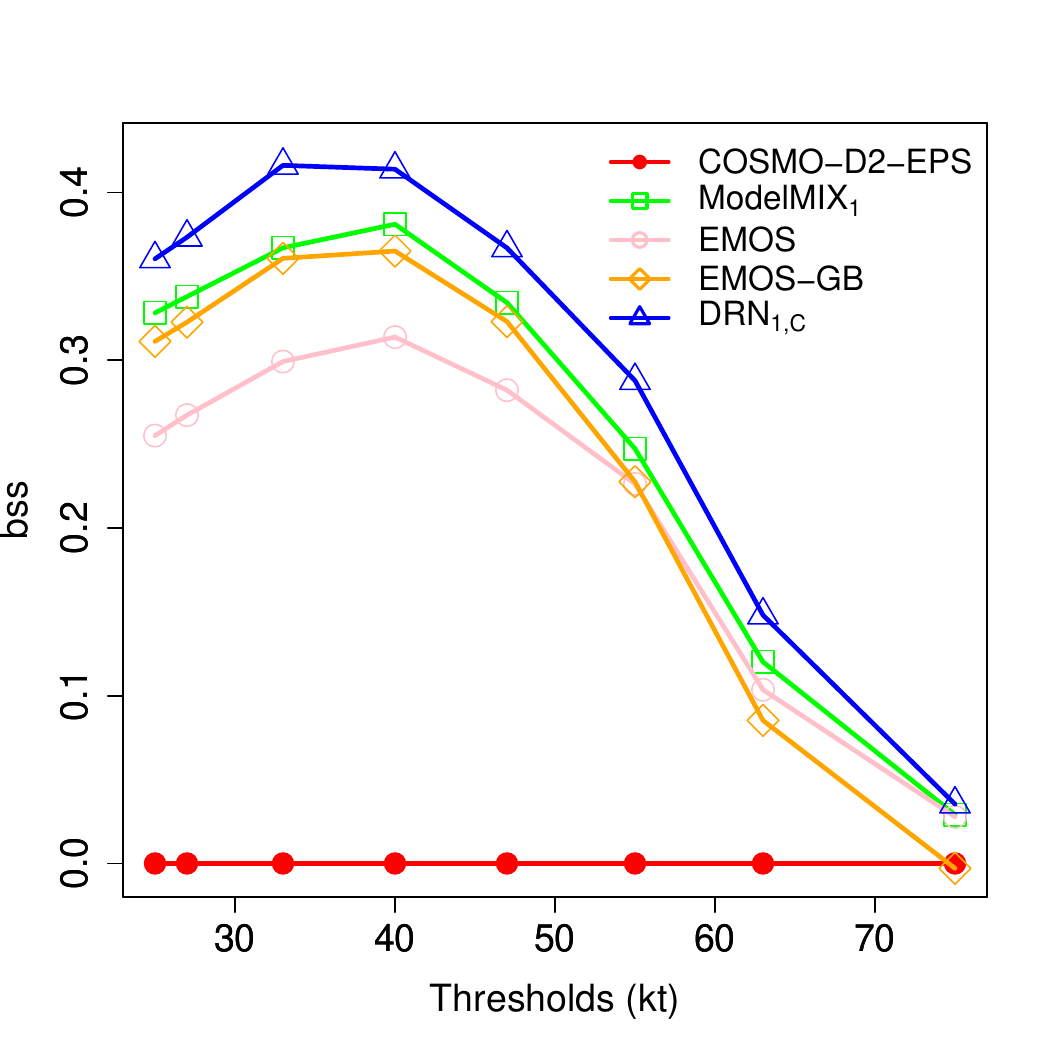}
\caption{Evolution of the BSSs of the postprocessing methods dependent on the wind threshold, where the reference is the BS of COSMO-D2-EPS. The evaluation is based on data from year 2020, all lead times, and the model run at 00UTC.}
\label{fig:Thresholds}       % Give a unique label
\end{figure}

Fig.\ \ref{fig:Thresholds} shows the BSS evolution with respect to the wind threshold for the model run initialized at 00UTC. Note that even though the BS depends on the frequency of the event, so the lower this frequency, the lower its value, and therefore both $BS$ and $BS_{ref}$ will by definition tend to zero for extreme events, the BSS is based on the division of both quantities so it behaves differently. According to the figure, higher differences are expected to happen for lower thresholds. 
For all methods, the improvements over the direct model output decrease for larger threshold values, but the ranking between the postprocessing methods mostly remains consistent. In line with the results of \citet{Schulz2022}, we find that the NN methods provide the best forecasts also for higher threshold values.
However, the evaluation might be less conclusive for very high threshold values due to the low event frequency, which implies large uncertainties in the score values, and the forecast distribution assigning smaller probability to this region will be preferred \citep[for details, see][Section 3.4]{LerchEtAl2017}.
Further, a detailed evaluation of individual case studies such as selected winter storms \citep{PantillonEtAl2018} could potentially highlight additional advantages or disadvantages of the different postprocessing methods, but is beyond the scope of this paper.

\subsection{Influence of NWP Model Changes}
\label{subsec:52}

The NN methods considered in the previous section were trained with the complete available dataset, i.e., data from 2010 to 2019. 
However, in March 2017 a major improvement occurred in the data assimilation method with the introduction of the KENDA technique. 
To evaluate the effect of these changes on the performance of the NN methods, we trained the same DRN model using two different training periods: from 2010 up to 2019, and only from March 2017 up to 2019. 
Note that the models include binary predictors to indicate the introduction of KENDA and COSMO-D2.
These networks were trained without persistence predictors, so they are also compared to a DRN trained using the complete dataset with persistence predictors. 
These three NN methods are used to produce forecasts for the year 2020. 

Further, in February 2021, the NWP model changed, as ICON-D2 replaced COSMO-D2. This was a major change in the NWP setup, so we have added another comparison with two more DRN variants, 
one trained with data from February 2021 up to June 2022, i.e., only ICON data, and another trained with data from March 2017 up to June 2022 (after KENDA was introduced). 
These models forecast the one year period from July 2022 to June 2023. 

Note that the frequency of the event differs from one verification year to the other. The uncertainty term of the BS is directly affected by the observed event, so changes in the frequency imply changes in the BS due to the contribution of the uncertainty term, even though the performance of the model might not change. Therefore, the verification results for both periods (2020 and 2022/2023) should not be directly compared.

\begin{figure}
\includegraphics[width=\textwidth]{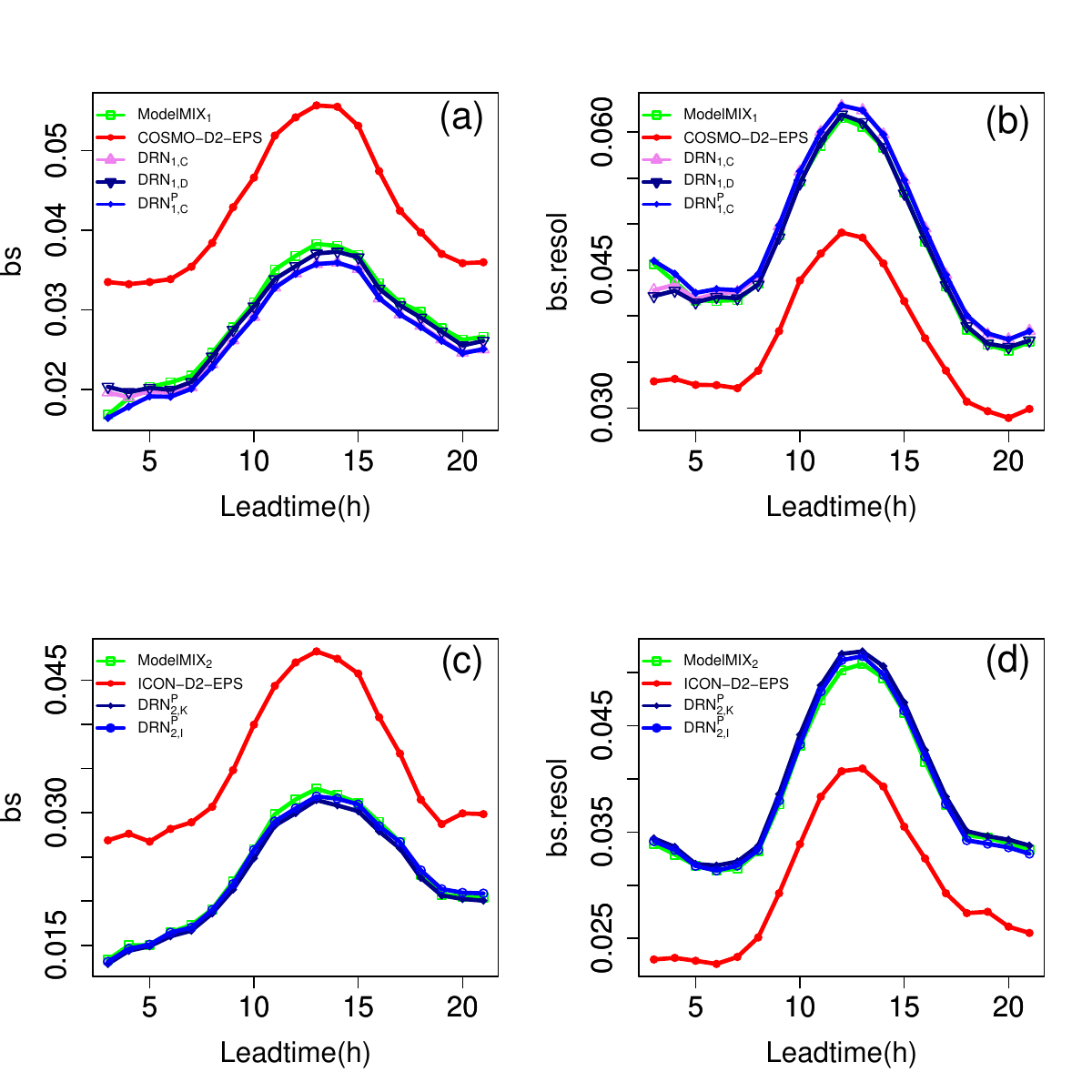}
\caption{Relevance of changes in the training period, due to: (a,b) changes in the assimilation process (KENDA; verified year: 2020) and (c,d) changes in the NWP model (ICON; verified year: 2022/2023). (a,c) BS; (b,d) resolution. The evaluation is based on wind threshold 25kt, and the model run at 00UTC.}
\label{fig:Training}  % Give a unique label
\end{figure}

Fig.\ \ref{fig:Training} shows the BS (a,c) and its resolution term (b,d) of the direct output of the NWP model and two postprocessing techniques, ModelMIX and DRN. The first row shows the verification for the year 2020 and compares performance when changes in the assimilation process (KENDA) were introduced. The second row shows the corresponding verification for the years 2022/2023 and compares two different underlying NWP models, COSMO and ICON, used in the training period. The reliability diagrams are similar to previous results, showing good calibration of the forecasts and thus are omitted. Major changes occur in the accuracy during the first hours whenever persistence predictors are used. These changes are more relevant than the training period length. Generally, longer training periods lead to slightly better results. For example, the BS improves mainly due to an improvement of the resolution, especially when forecasting midday. The difference in the BSs between the year 2020 and 2022/2023 are mainly due to the differences in the uncertainty term, i.e., due to the inter-annual variability of the observed frequency of the event.
In a nutshell, adding binary variables indicating the model version seems to work well, and the longer the training period, the better the results.

\subsection{Temporal Composition of Training Datasets}
\label{subsec:53}

When postprocessing ensemble forecasts in an operational context at large scales, computational complexity and technical requirements are more critical than in research studies. Hence, model configurations should aim at providing sufficiently good forecasts while keeping the computational complexity as low as possible. Due to the flexibility of NNs, several design choices can be considered when building the model.

One choice is whether to build joint models (over multiple spatial locations, and in a temporal context, i.e., over multiple lead times and/or model runs), or to use separate models in each individual case. 
While it has already been shown that building one NN model for all stations is feasible and improves the predictive performance, we briefly investigate how a joint model for all lead times and model runs compares to the previous practice of building separate models. As mentioned in Section \ref{ssubsec:pp_drn}, such a joint model increases the training set size by a factor of 38 in our case and vice versa reduces the number of model instances that need to be estimated by the same factor. This trade-off is then connected with the question of the length of the training period. 
Individual models have less data available for the same period of time and might thus require training periods that range back longer in time, and vice versa.
Here, we therefore compare NNs trained for each lead time and model runs with NNs that are trained jointly for all lead times and runs.
An added benefit of joint models is that fewer individual models need to be tuned and optimized. In our case, however, this is not relevant as we use the same architecture without any additional tuning for all variants.

\begin{figure}[t]
\includegraphics[width=\textwidth]{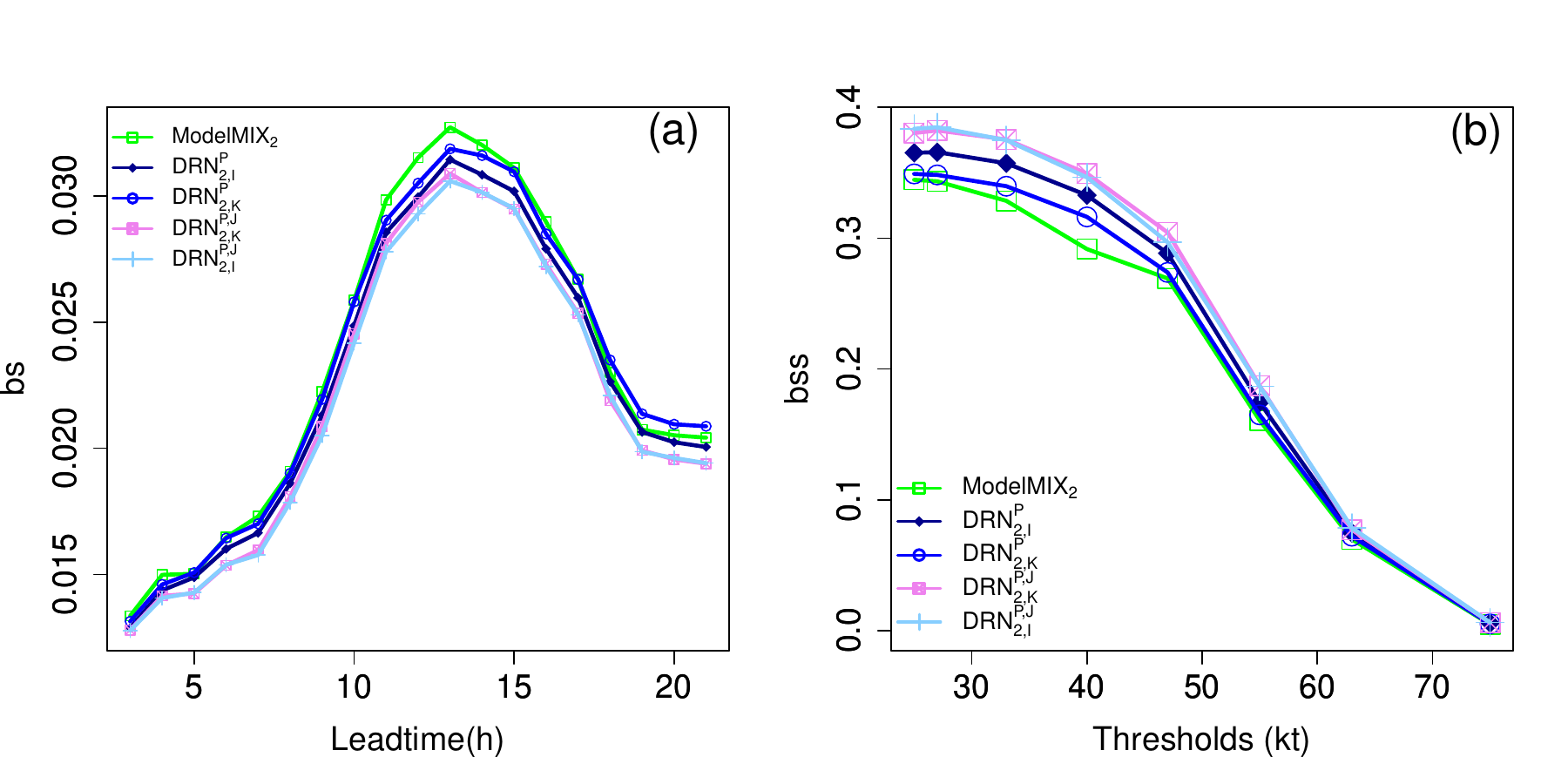}
\caption{Comparison of two NNs trained with all lead times and runs (indicated by a J in the exponent) and two NNs trained for each lead time and run. The evaluation is based on data from year 2022/2023 and the model run at 00UTC. (a) BS for wind threshold 25kt, (b) BSS for all lead times.}
\label{fig:Joint}  % Give a unique label
\end{figure}

Fig.\ \ref{fig:Joint} compares two DRN variants trained separately for each lead time and run for the period from March 2017 to June 2022 (DRN$_{2,\text{K}}^\text{P}$) and from February 2021 to June 2022 (DRN$_{2,\text{I}}^\text{P}$) with two networks trained jointly for all lead times and the two runs from 00UTC and 12UTC, one for the period from March 2017 to June 2022 (DRN$_{2, \text{D}}^\text{P,J}$) and another from February 2021 to June 2022 (DRN$_{2, \text{I}}^\text{P,J}$). All networks are trained with persistence predictors and forecast hourly probabilities of wind gust above 25kt for the NWP model run initialized at 00UTC from July 2022 to June 2023. All models are well calibrated. This increase of the training data set size does only lead to a minor improvement in performance, hence we argue that the additional costs for the more complex training scheme do not weigh up the improvements. Overall, we can conclude that more data leads to better results, but when a sufficient amount of training data is available, no more training data is beneficial in an operational context.
That said, further improvements might be possible with specifically targeted hyperparameter tuning for those settings. 

\subsection{Score Cards}
\label{subsec:Cards}
To support the decision process of implementing changes in the operational postprocessing of a meteorological center, it is useful to have diagrams summarizing the relevant verification results, for example showing traffic light colors (green if there is an improvement and red if there is none), that compare the new method with the method considered as reference. In this study, we have analyzed the benefit of applying postprocessing methods compared to the direct model output of the NWP system, and we have compared the performance of different postprocessing methods, namely EMOS and NN. Therefore, we utilize \textit{score cards}, i.e., diagrams showing the improvement by using different postprocessing techniques compared to our reference model, COSMO-D2-EPS for year 2020 and ICON-EPS for year 2023.

\begin{figure}
\includegraphics[width=\textwidth]{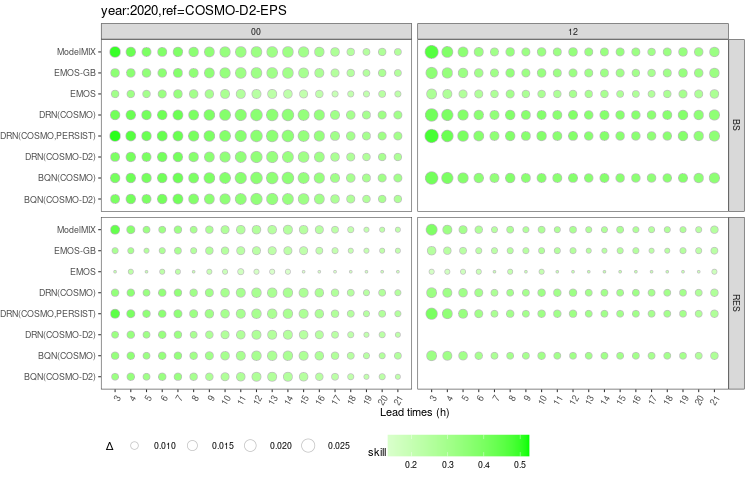}
\caption{Score cards comparing the benefit in BS and resolution of different postprocessing methods for wind threshold 25kt, year 2020 and two model runs (00UTC and 12UTC). Green color indicates an improvement after the postprocessing compared to the reference model, COSMO-D2-EPS. Size and color scale show how big the improvement is.}
\label{fig:Card2020}  % Give a unique label
\end{figure}

\begin{figure}
\includegraphics[width=\textwidth]{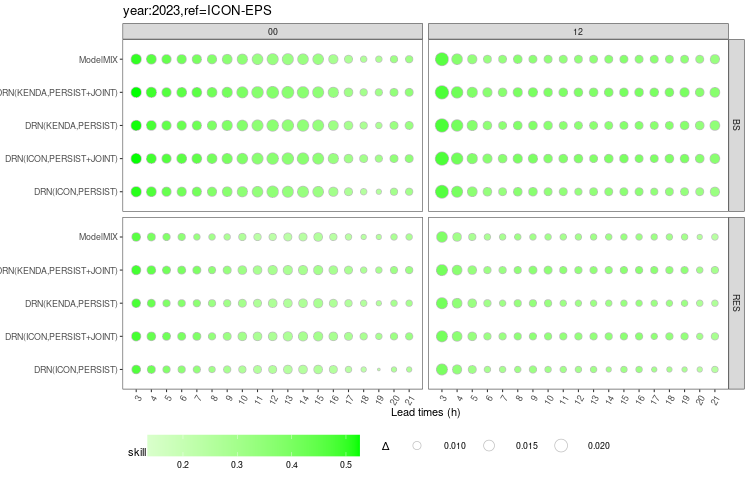}
\caption{Score cards comparing the benefit in BS and resolution of different postprocessing methods for wind threshold 25kt, year 2023 and two model runs (00UTC and 12UTC). Green color indicates an improvement after the postprocessing compared to the reference model, ICON-EPS. Size and color scale show how big the improvement is.}
\label{fig:Card2023}  % Give a unique label
\end{figure}

Results in this work show that all postprocessing methods always improve the direct model output and the performance of the different postprocessing methods mainly differs in the BS due to changes in the resolution. The temporal evolution of the differences in BS and resolution compared to the reference model are summarized in Fig.\ \ref{fig:Card2020} and \ref{fig:Card2023} for two years (2020 and 2023) and two runs (00UTC and 12UTC). The point size indicates the difference between the score of the postprocessing method and the score of the corresponding reference forecast. The color scale represents the skill of the score,
% that is, that difference divided by the value of the score of the reference model. 
as described in Eqs.\ \ref{eq:bss} and \eqref{eq:res_skill}. 

The benefit of applying ModelMIX or NN is much bigger than using EMOS or EMOS-GB. However, the difference between ModelMIX and NN scores are not relevant compared to the differences with the reference, and the circle sizes of ModelMIX and NN thus are similar for a particular lead time. The improvement is affected by the daily cycle, since the biggest circles for run 00UTC are around lead time 14h and smaller for the later lead times, whereas for run 12UTC bigger circles are shown for the earlier lead times and the differences are not much smaller for the later lead times.
Note that a circle in case of the resolution term is smaller than its counterpart in case of the BS, as the reliability term is additionally present in the BS difference (and becomes smaller by postprocessing).

\section{Summary and conclusions} 
\label{sec:conclusions}

We compare different techniques to postprocess the probability of hourly wind gusts exceedance, namely ensemble-MOS techniques and NNs. The classical EMOS and EMOS with gradient boosting are compared to the EMOS technique applied at the German weather service (DWD), ModelMIX, the distributional regression network (DRN) and the Bernstein quantile network (BQN). All methods are applied to the operational NWP ensemble prediction system of DWD, COSMO-D2-EPS (and later ICON-D2-EPS), over 170 SYNOP stations in Germany, for two years (2020 and 2022/2023). 

Forecast performance is measured in terms of calibration, sharpness and accuracy. Overall, we find that ModelMIX outperforms classical EMOS techniques, and NNs outperform ModelMIX. Both NN methods perform similarly. On the one hand, ModelMIX has the advantage of using all possible and more recent observations, and it contains persistence predictors. This leads to a better performance, in particular during the first hours. We further investigate the relevance of including persistence predictors in the NNs, showing that improvements over the first hours can be achieved this way, but there is no effect on later lead times. 
On the other hand, NNs predict a forecast distribution (instead of just the probability of exceeding a threshold), therefore, by default yield consistent probability forecasts at arbitrary thresholds. This work utilizes score cards summarizing the verification results to help in the decision process of changing the operational postprocessing technique.

In operational centers, improvements in the numerical model occur frequently, which leads to model changes during the training period of a NN. This issue is investigated, and forecasts produced by NNs trained on data including no model change are compared to NNs trained on data including relevant model changes. In particular, changes in data assimilation approach (KENDA) and the new numerical weather prediction model of the German weather service (ICON) are analyzed. Results show that the longer data sets are beneficial even if there are model changes. 
We here use binary variables to account for model changes. More sophisticated approaches such as transfer learning \citep{transfer_learning} might offer superior alternatives for the NN-based methods, and would be an interesting starting point for future comparisons.

Another important aspect for the operational use of postprocessing models at weather centers is the time efficiency (systems need to have a limited complexity to produce forecasts on time) and time consistency (forecasts should not jump from one time step to the next one or from one model run to the next one). 
While models estimated jointly over all lead times and model runs might have advantages in this regard, a detailed evaluation of potential forecast jumps is required to assess this aspect, and represents an interesting avenue for future research.

While we focused on the use of operational forecast datasets, some meteorological centers produce reforecasts by (re-)running the current model version for past forecast cases. This generates a larger training archive of cases with the current model setup, however, comes at high computational costs. In light of the improvements obtained via longer training datasets it would be interesting to investigate the effects of using reforecast data (without model changes) versus forecast data (with model changes) in more detail. From an operational perspective, this aspect is particularly relevant since the computational costs of generating reforecasts need to be carefully balanced against the costs of improvements to the NWP system (such as increases in the spatial or temporal resolution, or the number of ensemble members), see \citet{VannitsemEtAl2021} for a more detailed discussion of this aspect.

Evidently, our study only constitutes a first step and more detailed comparisons are needed in future research to comprehensively assess the respective advantages and disadvantages of the various postprocessing approaches. 
In particular, with a potential operational use of these methods in mind, several other aspects are of importance beyond the predictive performance assessed via standard verification metrics.
For example, ModelMIX is already applied to postprocess many other variables. While DRN has the advantage of providing full distribution forecasts, this comes at the cost of having to adapt the model for different target variables, e.g., by choosing an appropriate parametric forecast distribution. 
On the other hand, it might be possible to further improve the predictive performance of the NN models by tailoring them better to the task at hand, for example by moving towards categorical forecasting or classification models with suitably adapted loss function \citep[see, e.g.,][]{Baran2021}.
Finally, our study further illustrates that fair and direct comparisons between operational and new postprocessing techniques are challenging due to differences in terms of the specific setup, available inputs, and training dataset choices.
We here focused on the ModelMIX approach since it was possible to make the setups more comparable, even though ModelMIX does not run operationally at DWD. Instead, the operational technique is based on MOS applied to the global model (instead of to the limited area model) and in its deterministic version (instead of the EPS). 
% A comprehensive comparison to the operational system, ideally based on a suitable benchmark dataset or case study, would be an interesting avenue for future comparisons.

\section*{Acknowledgements}

Benedikt Schulz and Sebastian Lerch gratefully acknowledge funding within the project C5 ``Dynamical feature-based ensemble postprocessing of wind gusts within European winter storms'' of the Transregional Collaborative Research Center SFB/TRR 165 ``Waves to Weather'' funded by the German Research Foundation (DFG). Sebastian Lerch further gratefully acknowledges support by the Vector Stiftung through the Young Investigator Group ``Artificial Intelligence for Probabilistic Weather Forecasting.''
We thank Sebastian Trepte for providing the observation data, and Robert Redl for assisting in data handling.

%

% \clearpage
\bibliographystyle{myims2}
\bibliography{lib}

\end{document}